An internal disulfide locks a misfolded aggregation-prone intermediate in cataract-linked mutants of human γD-crystallin

**Eugene Serebryany*[1], Jaie C. Woodard*[2], Bharat V. Adkar[2], Mohammed Shabab[1], Jonathan A. King†[1], and Eugene I. Shakhnovich†[2]**

[1]*Department of Biology, Massachusetts Institute of Technology, Cambridge, Massachusetts 02139*
[2]*Department of Chemistry and Chemical Biology, Harvard University, Cambridge, Massachusetts 02138*
*\*Equal contribution.*
*†To whom correspondence should be addressed: jaking@mit.edu; shakhnovich@chemistry.harvard.edu*

Running title: *γD-crystallin aggregation precursor*

**Considerable mechanistic insight has been gained into amyloid aggregation; however, a large class of non-amyloid protein aggregates are considered "amorphous," and in most cases little is known about their mechanisms. Amorphous aggregation of γ-crystallins in the eye lens causes a widespread disease of aging, cataract. We combined simulations and experiments to study the mechanism of aggregation of two γD-crystallin mutants, W42R and W42Q – the former a congenital cataract mutation, and the latter a mimic of age-related oxidative damage. We found that formation of an internal disulfide was necessary and sufficient for aggregation under physiological conditions. Two-chain all-atom simulations predicted that one non-native disulfide in particular, between Cys32 and Cys41, was likely to stabilize an unfolding intermediate prone to intermolecular interactions. Mass spectrometry and mutagenesis experiments confirmed the presence of this bond in the aggregates and its necessity for oxidative aggregation under physiological conditions *in vitro*. Mining the simulation data linked formation of this disulfide to extrusion of the N-terminal β-hairpin and rearrangement of the native β-sheet topology. Specific binding between the extruded hairpin and a distal β-sheet, in an intermolecular chain reaction similar to domain swapping, is the most probable mechanism of aggregate propagation.**

Partially unfolded or misfolded, aggregation-prone protein conformational states are linked to a wide array of age-related protein misfolding diseases. The best studied of these conditions include amyotrophic lateral sclerosis (superoxide dismutase), Parkinson's disease (α-synuclein), serpinopathies ($α_1$-antitrypsin), cancers with P53 and P21 tumor suppressor defects, and lens cataract (crystallins) (1-5). Some of these aggregates contain the well-known amyloid structure, and others do not; even in cases where amyloid is the final aggregated state, oligomers, prefibrillar species, and amorphous aggregates are often closely linked to pathology (6,7). Structures and interactions of specific locally unfolded or misfolded intermediate states are critically important in the mechanisms of non-amyloid aggregation (1). Here we investigate such an intermediate-based aggregation mechanism for a lens γ-crystallin.

Since the critical properties of protein structure – thermodynamic and kinetic stability, as well as alternative, misfolded, or partially unfolded conformational states – are encoded ultimately in the sequence, point mutations have been found to underlie familial forms of virtually all misfolding diseases. However, most conformational diseases are sporadic in origin, with neither inherited nor *de novo* mutations present in the relevant genes. In the absence of mutations, such cases may arise from direct chemical modifications of residues within the relevant polypeptide. Well-studied examples include superoxide dismutase 1, the eye lens crystallins, and α-synuclein (4,8-10). In a great number of other cases, post-translational modifications cause conformational change as part of the normal function of a protein. During the course of aging, both somatic mutations and side-chain modifications tend to accumulate, further increasing the heterogeneity of the proteome (11,12).

The human lens grows slowly in layers throughout life; its core, however, is metabolically inert and therefore provides a striking case study in protein aging (13). Lens core cells are enucleated, contain no organelles or protein synthesis and diminished degradation machinery, and are never replaced (13-15). Crystallins comprise over 90% of their total protein content. Lens transparency depends on the crystallins' structural integrity and lack of long-range packing even at very high concentrations (16). Passive chaperones, the α-crystallins, are present in lens, but their capacity declines with age, even as their substrates, the βγ-crystallins, accumulate destabilizing chemical modifications (13) due to a variety of environmental damage (17). The result is generation of partially unfolded intermediate conformational states and progressive increase in light scattering (lens turbidity) due to aggregation (18-21). No long-range structure, amyloid or otherwise, has been found in the cataractous aggregates (22,23), except in certain rare congenital cases (24,25), but disulfide bonds and other covalent modifications are common (13,26-29).

Non-native disulfides can act as kinetic traps in protein folding and misfolding (30-33). Hence, some pathways of protein aggregation depend strongly on the redox environment. Notable examples include superoxide dismutase 1 (34,35) and $β_2$-microglobulin (36-38). We and others have demonstrated that γ-crystallins, too, form crucial disulfide-mediated interactions. Thus, aggregation of W42Q γD-crystallin depends on internal disulfide bonds (39), and so do the folding kinetics and protease sensitivity of γB-crystallin (40). Although the crystallins are cytosolic, and therefore initially fold in a reducing environment, the redox potential of lens cytosol shifts toward increasingly oxidizing values during the course of aging and, especially, cataractogenesis (41). Accordingly, the proportion of Cys residues forming disulfide bonds increases over time (42), and mature cataracts show disulfide cross-linking in over half the lens protein content (43).

To better understand the nature of the aggregation precursor and the mechanism of non-amyloid aggregation for human γD-crystallin (HγD), we iterated between experiments and simulations in studying the oxidation-mimicking W42Q mutant as well as the W42R congenital-cataract mutant. Despite the difference in charge, the two mutants' thermostabilities and aggregation properties were nearly identical, and both required formation of an internal disulfide bond for aggregation under phyiological conditions *in vitro*. All-atom, two-chain tethered Monte-Carlo simulations using a knowledge-based potential predicted that, of the six possible internal disulfides within the mutated N-terminal domain, the Cys32-Cys41 bond in particular trapped a partially misfolded state prone to specific intermolecular interactions. Experiments confirmed this bond was indeed required for aggregation *in vitro*. Independently, a recent proteomic study has linked oxidation of these residues to age-onset cataract *in vivo* (44). Constraining the simulations to the now-established disulfide enabled us to propose the most likely structure of the locally unfolded aggregation precursor, as well as a polymerization mechanism that shares key features with domain swapping.

**EXPERIMENTAL PROCEDURES**

*Protein expression and purification*

All proteins were expressed and purified as described (39). Briefly, BL21-RIL *E. coli* (Agilent) were transformed with pET16b plasmids containing the relevant cDNAs under the T7 promoter with no epitope tags. Cells were induced at 18 °C overnight. Cell pellets were stored at -80 °C. Ammonium sulfate was added slowly to cleared cell lysate to ~30% to precipitate lowly-soluble contaminants, then crystallins were precipitated by increasing ammonium sulfate concentration to ~50%. The pelleted proteins were washed and further purified by size-exclusion chromatography. Purity was >95% by SDS-PAGE. Typical yields for the W42Q and W42R mutants were 30-40 mg / L of cell culture in Super Broth, 10-15 mg/L for C18A/W42Q, and 50-70 mg/L for C32V/W42Q and WT proteins. Human γS-crystallin was purified as described (45).

*Differential scanning calorimetry*

Thermal denaturation of WT and mutant proteins was carried out by differential scanning calorimetry (nanoDSC, TA instruments) using 0.3 or 0.6 mg of protein. Prior to the experiment, proteins were reduced with TCEP for 1-2 h at room temperature, then dialyzed three times against DSC buffer (10 mM sodium phosphates, 50 mM NaCl, pH 7). TCEP to a final concentration of 0.5 mM was added to the samples immediately prior to DSC. The buffer from dialysis reservoir was used as reference for the DSC experiments. The samples were heated from 25 to 90 °C at a scan rate of 1 °C/h. Thermodynamic parameters were derived by fitting the data to a sum of two 2-state unfolding models, one for each domain, using NanoAnalyze (TA instruments).

*Aggregation assays*

Aggregation of purified samples was induced by shifting to 37 °C or as indicated in sample buffer (10 mM ammonium acetate, 50 mM NaCl, pH 7) with 1 mM EDTA. Turbidity was monitored using a FluoStar Optima plate reader (BMG) and polypropylene 96-well plates (Greiner) in 200 µl volume with no agitation. In the case of Figure 1b and Figure 2, the measurements were performed in a quartz cuvette in a spectrophotometer. Oxidation-dependent aggregation experiments were performed in the same way but with controlled degree of oxidation.

Purified protein samples were reduced by incubation with 0.5 mM TCEP for 30-60 min. at room temperature with gentle agitation. The reducing agent was removed by 2-3 rounds of dialysis against 4 L of sample buffer, one of which was overnight. Degree of oxidation (OxD) was set using mixtures of reduced and oxidized glutathione (GSH and GSSG) at a total concentration of 2 mM (counting GSSG as two molecules), according to the formula:

$$OxD = \frac{2[GSSG]}{[GSH]+2[GSSG]}$$

*Fluorescence*

Protein samples (4 µM, in 10 mM ammonium acetate buffer, 50 mM NaCl, 1 mM EDTA, pH 7) were equilibrated in heated cuvettes at the indicated temperatures for 10 min prior to the experiment. Intrinsic tryptophan fluorescence spectra were obtained using a Cary Eclipse fluorimeter (Varian) with excitation at 280 nm and emission scanned between 300 and 400 nm to make sure the signal reached baseline by 400 nm.

*Monte-Carlo unfolding simulations*

All-atom (non-hydrogen) Monte-Carlo simulations were performed using a program described in previous publications, developed by the Shakhnovich group (46-49). The program incorporates a knowledge-based potential, with terms for contact energy, hydrogen-bonding, torsional angle, and sidechain torsional terms, as well as a term describing relative orientations of aromatic residues. The move set consists of rotations about backbone and sidechain torsional angles, with bonds and angles held fixed. Software for the Monte Carlo program is available on the Shakhnovich group website: http://faculty.chemistry.harvard.edu/shakhnovich/software.

Human γD crystallin (PDB ID 1HK0) was used as the initial structure in simulations. Mutations (W42Q, W42R, T4P, W130E) were introduced using PyMOL. For WT and each mutant, an initial simulation was run at low temperature (T = 0.150 in simulation units) for 2,000,000 steps. Starting from the final structure of this simulation, a 60,000,000 step simulation was run, for each of 32 temperatures, ranging from T = 0.1 to 3.2. The final frame was extracted from each of 50 separate runs, and the average RMSD from the native structure was calculated for the N-terminal and C-terminal domains separately. Curves representing the dependence of RMSD from the native structure on temperature were fit to a sigmoidal function; since the domains were fitted separately, the sigmoid corresponded to a two-state model of unfolding. Representative structures were obtained for W42R by clustering contact maps (residue-residue α-carbon distance less than 10 Å) for final simulation frames, using the MATLAB clusterdata function with a cutoff of 0.9. Images were generated for representative structures from the two most populated clusters. A further 300 simulations at T = 0.8 were then conducted.

*Two-chain tethered unfolding simulations*

Two-molecule single chain simulations were carried out for two γ-crystallin molecules connected by a 12-residue linker. The amino acid sequence of the linker was GSGSGSGSGSGS, and the initial structure is shown in **Figure 5A**. Each simulation was 80,000,000 Monte Carlo steps in length. 300 simulations were run for each of WT, W42R, and W42Q. In addition, simulations were run with a harmonic constraint applied to each pair of disulfides within the N-terminal domain, holding sulfur atoms of the cysteine residues approximately 2 Å apart. The same potential maintained the disulfide once formed. Cys-Cys angles were neglected for these simulations, unless disallowed by steric effects.

Simulations were mined for protein-protein interactions every 5,000,000 frames. Residues were said to be in contact if their α-carbons were within 10 Å of each other. A protein-protein interaction was counted when more than 50 pairwise residue-residue contacts were detected between proteins. An interaction was said to be an antiparallel β-strand interaction if it included more than 6 residues in a row in an antiparallel configuration. An interaction was said to be a parallel β-strand interaction if the interaction included more than 6 residues in a row in a parallel configuration. The interaction was said to be native-like if greater than 10 of the interactions were present between domains of the native structure (see, for instance **Figure 5C**). For parallel and antiparallel β-strand interactions, the β-strands involved in interaction were recorded, according to the numbering system in **Figure 7**.

*Disulfide mapping*

The predominant disulfide bond was mapped in W42Q aggregates produced upon incubation at 37 °C and OxD=0.2 for ~2 h in sample buffer with 1 mM EDTA. Aggregates were pelleted by centrifugation of the turbid mixture at 14,500 x g for 5-10 min. The pelleted fraction was washed three times with sample buffer, then resuspended in 0.1 M MES buffer, pH 5.5. The resuspended sample was kept at 37 °C for 20-30 min to allow dissociation of weakly bound protein, leaving stable aggregates. These aggregates were solubilized by addition of guanidine hydrochloride to a final concentration of 4 M and denatured by incubating at 37 °C for 3 h, at pH 5 and in the presence of 1 mM EDTA. The low pH and chelator were used in order to inhibit Cys oxidation and disulfide scrambling during denaturation. Denatured samples were refolded by gradual dilution of the guanidine-containing buffer with dilution buffer (0.1 M MES, pH 5.5, 1 mM EDTA) at the rate of 25% dilution every 15 min. Once [Gdn] had been brought down to 2.6 M, samples were centrifuged and concentration determined by Nanodrop (Thermo Scientific). Dilution buffer was then added to reduce [Gdn] to 2.0 M, and then TCPK-treated bovine trypsin (Sigma-Aldrich) was added at the ratio of 1:20 w/w. Dilution continued until [Gdn] fell to 1.0 M; samples were centrifuged again and concentration determined; and additional trypsin was added to bring the total trypsin:protein ratio to 1:10. Samples were then incubated for 12-16 h at 37 °C to allow digestion.

Afterward, digests were centrifuged to remove any precipitated material and trypsin activity quenched by addition of trifluoroacetic acid to 0.3% v/v final concentration. The high trypsin:protein ratio was necessitated by the partly denaturing, non-reduced digestion conditions. Stock (pre-reduced) W42Q sample was used as a control for the denaturation and digestion steps to check for adventitious disulfides that might arise during sample handling. We used W42Q in preference to W42R in order to facilitate isolation of Cys41-containing peptides from the tryptic digests.

Tryptic digests were separated on a 5 μm, 250 x 4.6 mm C18 column (Higgins Analytics) on an Agilent 1200 HPLC instrument using a 36 min. 5-60% gradient of acetonitrile with 0.1 % trifluoroacetic acid. Absorbance at 280 nm was monitored in order to enable quantitative comparisons of peak intensity independent of their ionization efficiency. Peaks of interest were collected manually and identified separately by brief HPLC over a 2.7 μm C18 column (Agilent) coupled to an Agilent 6230 electrospray ionization mass spectrometer. Multiply charged states were identified and deconvoluted manually.

**RESULTS**

*W42R and W42Q are similar to each other in stability and aggregation properties*

Trp 42, located in the bottom of the N-terminal domain core (**Figure 1A**), is particularly sensitive to substitution, as both oxidation-mimicking W42E and W42Q substitutions *in vitro* and the congenital cataract-linked W42R substitution destabilize the protein and lead to aggregation (39,50,51). The congenital and oxidation-mimicking substitutions differ in their charge. Nonetheless, the W42R sample showed very similar aggregation properties to W42Q. The purified protein was stable for weeks in storage at 4 °C, but aggregated in a redox-dependent manner when shifted to a higher temperature (**Figure 1B**). Accommodation of charged residues in protein cores has been the subject of intensive study (52,53), thus it was of interest to compare the charged (W42R) and neutral (W42Q) core substitutions. DSC (**Figure 1C**) revealed that there was almost no difference in the temperature of the two mutants' unfolding transitions. The available chemical denaturation curves confirm similar stability of W42Q and W42R (39,51). **Table 1** summarizes melting temperatures obtained from DSC of the WT and several mutant constructs.

*W42R and W42Q aggregation occurs at solution redox potentials found in cataractous lens*

Disulfides are known to form in lens cytosol during the course of aging and cataractogenesis as the pool of lens glutathione becomes smaller and more oxidized over time (9,28,41,44). To mimic this physiological oxidation process *in vitro*, we carried out aggregation experiments by first reducing the proteins completely and then placing them at a series of known redox potentials using glutathione buffers. In those experiments, both mutants showed highly oxidation-dependent aggregation at physiological temperature and pH. The range of aggregation-permissive OxD (oxidation degree, defined as the ratio of oxidized/total glutathione) values was comparable to that found in the lens during cataractogenesis (41) (**Figure 1D**). While both mutants showed clear aggregation by OxD = 0.18, the WT protein did not show any oxidation-dependent aggregation even at OxD = 0.5, suggesting that aggregation requires a partially unfolded or misfolded conformation accessible at physiological temperatures only as a result of the mutations or comparable damage (**Figure 1D**).

We have previously reported that the aggregates formed by the temperature-sensitive Trp mutants of γD-crystallin under near-physiological conditions are neither disulfide-bridged nor amyloid and expose few hydrophobic residues (39,50). These observations raised the question of whether a seeding effect exists in this aggregation pathway. Seeding is typical for templated aggregation phenomena, such as amyloidogenesis or crystallization: pre-formed aggregates added to a fresh solution of the protein serve as nuclei, bypassing the lag phase and greatly accelerating the aggregation process. However, as shown in **Figure 2**, this is not the case for W42R aggregation under physiological conditions. When a sample from the aggregating protein was mixed in with an identical fresh solution, turbidity initially decreased slightly, and the subsequent lag time was not significantly shorter than for to the initial aggregation. The conformational conversion in W42R HγD does not appear to be a templated one.

*Oxidation causes a conformational change in the mutant proteins*

Intrinsic tryptophan fluorescence is a highly sensitive reporter of conformational change in γD-crystallin, as evidenced by many previous studies (18,20,39,54,55). As shown in **Figure 3**, baseline (non-oxidized) fluorescence spectra show no conformational difference between the WT and mutant proteins. This is consistent with the data in **Figure 1C**, where unfolding is not detectable for any protein at 37 or 42 °C under reducing conditions. However, when the solution redox potential is oxidizing, the mutants display slightly but reproducibly red-shifted spectra (**Figure 3A, B**), in contrast with the WT protein (**Figure 3C**). We attribute this change to increased solvent exposure of at least one Trp residue in the mutant proteins, most likely Trp68, since that is the only Trp residue in the destabilized N-terminal domain of these mutants. This spectral change was observed over the course of 60 min; no red shift was observed when oxidizing agent was omitted (**Figure 3D**). Thus, aggregation of these mutant proteins correlates with a conformational change stabilized by oxidation.

*Simulation of oxidative misfolding reveals characteristic early intermediates*

Single chain simulations showed that cataract-associated mutations in the N-terminal domain destabilize the N-terminal domain (**Figure 4A**), while the oxidation-mimicking mutation W130E in the C-terminal domain destabilizes the C-terminal domain relative to WT (**Figure 4B**). In particular, W42Q and W42R both led to destabilization of the N-terminal domain, with very similar apparent melting curves, consistent with experiment (**Figure 1C**). Even in the WT protein, the N-terminal domain was less stable than the C-terminal domain, also consistent with past experimental results (18,55,56) and simulations (57). Also consistent with experiments, the W130E mutation largely eliminated the differential domain stability, thus leading to a more cooperative unfolding transition (50). Partially unfolded structures from W42R simulations at intermediate temperature show that separation of the two N-terminal Greek keys and detachment of the N-terminal hairpin are early events in the unfolding process (**Figure 3C**). As we discuss below, flexibility of the loop connecting the detached hairpin to the rest of the domain core may allow Cys residues that are distant in the native structure to come together and form a disulfide bond.

Of 300 independently simulated unfolding pathways, five were remarkable in that the N-terminal hairpin, having detached from its native position, annealed to another β-sheet on the molecule, extending it. The final states of three of these runs are shown in **Figure 5A**, and the other two in **Figure 5B**. In both cases, the observed internal topological rearrangements suggest possible binding sites for aggregate propagation.

*Two-chain tethered simulations predict internal-disulfide structures prone to extensive intermolecular interactions*

Whole-protein mass spectrometry experiments (39) revealed that most monomers within the aggregated fraction contain only one disulfide bond. HγD contains six Cys residues (**Figure 1A**), so we set out to determine whether a particular internal disulfide is most likely to generate intermolecular interactions that could lead to a polymeric state. This was accomplished by simulating two identical mutant chains tethered by a flexible (Gly-Ser)$_6$ linker. Disulfide bonding was simulated by placing a spring potential between pairs of thiols, leading to a preferred 2-Å final sulfur-sulfur distance. In all cases, the simulated internal disulfides formed prior to the intermolecular interactions (**Figure 6A**). Protein-protein interactions were found in a relatively small fraction of simulation frames for WT and mutant proteins, including W42R with internal disulfide bonds (**Figure 6B**). Interestingly, disulfides between cysteine residues that are adjacent in sequence led to a greater number of protein-protein interactions than disulfides between non-adjacent cysteines (**Figure 6B**), consistent with previous work on lattice proteins (58). The type of interaction (**Figure 6C-G**) depended on the particular mutant and disulfide bond formed. In particular, disulfide bonding between the second and third cysteines in the N-terminal domain led to the greatest number of intermolecular antiparallel β-sheet interactions. Therefore, this disulfide bond, between residues Cys32 and Cys41, was predicted as the likeliest to generate redox-sensitive aggregates.

*LC/MS and mutagenesis identify Cys32-Cys41 as the predominant disulfide crosslink responsible for aggregation*

We mapped the predominant disulfide in W42Q aggregates produced by mild oxidation (OxD = 0.2) using mass spectrometry. When samples were digested under non-reducing conditions, two new peaks appeared in the HPLC traces of the aggregated sample relative to the stock control (**Figure 7A**). The identities of select HPLC peaks were assigned by electrospray mass spectrometry, where disulfide-containing peptides are distinguished by a -2 Da. mass shift. These experiments revealed Cys32–Cys41 as the likely cross-link, which is in agreement with both our computational results and the most recent lens proteomic data (44). To detect the disulfide directly, we carried out tryptic digestion without a reduction step, as previously reported (59). A drawback of this approach is increased likelihood of missed tryptic cuts, as well as chymotryptic cleavages. However, these features also offered the advantage of multiple independent detections of disulfide-bonded peaks within the same experiment. Furthermore, tryptic digestion is known to be particularly inefficient within tight secondary structures such as β-turns. Accordingly, most of the digested samples had a missed cleavage between the bonded Cys residues, at Arg36. Digestion efficiency at this position was also higher in the absence of the disulfide bond (**Figure 7A**). Since missed cleavage at Arg36 makes Cys32 and Cys41 part of the same peptide, we were able to conclude that the disulfide identified here is, in fact, internal, rather than intermolecular. Since size-exclusion traces of the OxD=0.2 supernatant did not reveal any dimers, we conclude that intermolecular disulfides were unlikely to form under these conditions.

We further confirmed the disulfide mapping results using mutagenesis. Cys residues were mutated in the background of W42Q. In this background, Ala substitutions at Cys41 or Cys78 did not yield soluble protein, likely due to excessive loss of stability of the already destabilized mutant N-terminal domain. The C18A/W42Q double mutant could be purified, albeit with reduced yield. The C32A/W42Q construct's yield was impractically low, so we resorted instead to a Val substitution, which is present at this position in many related crystallin sequences. These two mutants are predicted to have divergent aggregation phenotypes: C18A/W42Q is predicted to retain its ability to aggregate by the pathway we have proposed, while C32V/W42Q is expected to lack this ability.

Oxidative aggregation was found to be highly temperature-dependent, consistent with a requirement of conformational change (39,50). Thus, W42Q showed no aggregation at 32 °C even under the most oxidizing conditions employed; oxidation-dependent aggregation at 37 °C; and strong aggregation even at low OxD values at 42 °C (**Figure 7B,C,D**). As shown in **Figure 7E**, the C32V/W42Q double mutant is ~5.6 °C more stable than W42Q, whereas C18A/W42Q is ~2.9 °C less stable. To control for the effect of thermostability, we carried out oxidative aggregation assays at 32 °C for C18A/W42Q and 42 °C for C32V/W42Q. In contrast to W42Q, C18A/W42Q aggregated in an oxidation-dependent manner already at 32 °C, while C32V/W42Q showed no aggregation even at 42 °C (**Figure 6F,G**).

*Molecules with Cys32-Cys41 internal disulfide are predicted to aggregate by β-sheet completion*

Having established experimentally by LC/MS, and confirmed by mutagenesis, that the Cys32-Cys41 internal disulfide was indeed required for aggregation, we carried out a more detailed analysis of two-chain simulations involving this disulfide bond. The observed parallel and antiparallel interactions were classified based on which β-strands were in contact. Most of these interactions were between strand 1 and strand 14, or between the first strands of the two proteins (**Figure 8**; see also **Figure 6E,F**). In the former case, the interaction could be perpetuated, as the N-terminal hairpin of each protein binds to the folded or mostly-folded domain of the next protein, extending a β-sheet that is formed in the native structure (model in **Figure 9**).

**DISCUSSION**

Age-onset cataract is a highly prevalent disease as well as a case study in protein aging. Its pathology has multiple contributing factors, including exposure to UV light, tobacco, and certain specific chemicals; however, the effect is generally the same – aggregation of the eye lens crystallins (5,13,16,17,60,61). In line with the multiple causality of the disease, several distinct biochemical perturbations *in vitro* have been shown to cause crystallin aggregation (54,62-64). Cataract-associated point mutations, especially ones mimicking *in vivo* chemical modifications, have been highly useful in dissecting the mechanism of aggregation (21,39,51) and developing interventions to modify it (65). Synergistic effects are likely among the various modes of modification or damage. Thus, our combined computational and experimental results suggest a link between Trp oxidation, often the result of UV exposure, and disulfide bonding, a marker of more general oxidative stress in the aging lens.

Our All-Atom Monte Carlo program has been used to successfully predict folded states and relative mutant stabilities of single proteins in previous publications (46,49). Here, we used the program to simulate interactions between two proteins by connecting them by a flexible linker. Such an approach is commonly used in experimental single-molecule pulling studies, including of HγD (66), and has been used in the past to computationally study protein-protein interactions (67,68). While we are currently developing a version of the Monte Carlo program with multiple freely diffusing chains, the peptide linker approach is a straightforward way to adapt a program designed for single chain simulations to study of the initial steps in protein aggregation. In fact, the approach yields reasonable results: the strands most often involved in protein-protein interaction are protected in the native structure but become exposed early in single-chain simulations. We expect that peptide-based drugs targeting these strands may slow or prevent aggregation of γ-crystallin molecules.

Using molecular dynamics simulations with subsequent annealing of two γ-crystallin molecules, Das *et al.* found interactions between the unfolded N-terminal domain and strands 13-15 (69). Similar regions interact in our simulations, but the interactions are more specific. Thus, we observe β-sheet completion, involving hydrogen bonding between β-strands. The speed of our program relative to conventional MD techniques allows for a greater sampling of configurational space necessary to obtain such structures, as well as obviates the need to simulate under highly denaturing conditions (such as the 8 M urea used by Das *et al.* (70)). We propose that aggregation proceeds from oligomerization in which the N-terminal hairpin binds to an exposed region of the C-terminal domain. Specifically, the N-terminal hairpin detaches from the N-terminal domain, and strand 1 forms antiparallel hydrogen bonding interactions with strand 14. While such a structure remains to be experimentally validated, it is consistent with the lower folding stability of the N-terminal domain relative to the C-terminal domain and the observation that many cataract-associated mutations in γ-crystallins cluster in or near the N-terminal hairpin (20). It also agrees with the recent report that the N-terminal β-hairpin is involved in domain swapping in HγD polyproteins under mechanical force (66).

That internal disulfide bonds stabilize specific partially unfolded states in HγD was not appreciated until recently (39). Although our statistical potential does not fully account for the distinct chemical properties (e.g., hydrophobcity) of disulfides and thiols, we have incorporated the geometric constraints on folding that result from disulfide formation. We found that disulfides between cysteines close together in sequence, particularly Cys32 and Cys41, stabilized a partially unfolded intermediate state prone to specific intermolecular interactions. It is worth noting that disulfide bonding is critical to the aggregation pathways of other well-known disease-related proteins, including superoxide dismutase 1 and β$_2$-microglobulin (34,35,37,38). Conversely, disulfide engineering has been used to trap important folding intermediates in a variety of proteins (71).

Surprisingly, among the possible disulfides in HγD, our study pointed to one in particular, Cys32-Cys41, as most likely to stabilize the aggregation precursor. Our results agree with a recent proteomic report linking the oxidation state of Cys32 and Cys41 in γD-crystallin to cataractous aggregation in human lens (44). Thus, it may be possible to reconstitute the *in vivo* cataractogenic biochemical pathway both *in vitro* and *in silico* to reveal structures of transient intermediates and specific misfolded states en route to "amorphous" aggregation. It is noteworthy that oxidation of Cys41 to a disulfide is one of the strongest correlates of human (but not murine) cataract development identified by Monnier and colleagues, and is found in both γD- and γC-crystallin (44). However, it has been unclear in the proteomic studies whether such disulfides are intra- or intermolecular. The aggregation in our *in vitro* assays is due to an internal bond, since both Cys residues are part of the same disulfide-containing peptide (**Figure 7A**). Studying aggregates by high-resolution structural techniques *in vitro* is very challenging, leading to a paucity of atomic-level mechanistic detail. Simulated structures based on the Cys32-Cys41 bridge provide a possible molecular mechanism for misfolding and polymerization into light-scattering aggregates (**Figure 9**).

The aggregation model proposed in **Figure 9** shares many features with classical domain swapping, as proposed by the Eisenberg group (72), yet the polymers do not reconstitute the native state of the monomeric protein. Horwich pointed out the possibility that domain swapping may reconstitute not the native state, but a partially unfolded or misfolded state that is at least kinetically stable (1). Since we have observed intramolecular strand 1 – strand 14 binding in single-chain simulations (**Figure 5**), this may be an example of such a mechanism. Stability of the internally rearranged monomer depends on the length and flexibility of the linker connecting it to the rest of the domain core. Our simulations have shown that this linker is sufficiently long and flexible in the absence of a disulfide bond. Shortening or straining a linker between two structural elements is the most common way to generate classical domain swapping (73,74). Formation of the Cys32-Cys41 disulfide, which we find to be required for polymerization, creates a loop within this linker sequence and may thus accomplish both. Reducing this disulfide bond is expected to make the monomeric state favorable again, consistent with our observation that addition of reducing agent can reverse the aggregation process (39).

We note that the internal rearrangement to form a "cyclic" monomer is expected to compete with the aggregation process, yet it is also likely to facilitate formation of the internal disulfide bond needed for aggregation. A potentially better way to inhibit aggregation would be to block the binding edge of strand 14 with a peptide or small-molecule inhibitor that would compete with the N-terminal β-hairpin. Finally, a major challenge for inhibition of classical domain swapping reactions is that any intervention to stabilize the native state is likely to also stabilize the polymer. If a partially misfolded state is the origin of the swapping, however, then inhibitors can be targeted to a specific part of the native structure.

The primary model of protein-protein interactions suggested by our simulations extends the self-recognition and domain-swapping paradigms currently associated with protein aggregation. Misfolding simulations of multidomain proteins such as titin suggest a role for domain swapping and/or self-similar amyloid-like interactions in misfolding and aggregation (75,76), and these studies have now been extended to HγD (66). In our predicted oligomerizing structure, strand 14 is sandwiched between two homologous β-strands, one belonging to the native structure (strand 13) and the other belonging to another domain or protein (strand 1). Thus, the non-native extended β-sheet structure is formed via native-like interactions that are not actually present in the native structure but that mimic native contacts on the other side of the β-strand. It will be interesting to see whether such "native-like" interactions between regions homologous to natively interacting ones are found in other aggregating proteins.

**ACKNOWLEDGEMENTS**

This work was supported by NIH grant R01 GM111955 to E. I. S.; NIH Biophysics training grant to J. C. W.; NIH grant R01 EY015834 to J. A. K.; and fellowships from MIT Biology and the Whitehead Institute for Biomedical Research to E. S. We are grateful to Prof. Elizabeth Nolan (MIT) for advice on experimental design; Cameron Haase-Pettingell for protein expression; to Phoom Chairatana and Lisa S. Cunden for training on HPLC and ESI-MS; and to Dr. Catherine Goodman for editorial suggestions.

**FIGURE LEGENDS**

**Figure 1: Aggregation of W42Q and W42R human γD-crystallin is highly redox-dependent.** (A) Backbone ribbon diagram of the native, wild-type γD structure (PDB ID 1HK0) showing the Trp residues as space-filling models and Cys residues in stick representation. Trp 42 is colored red. (B) Turbidity traces of W42R at 150 µM upon shifting

from 4 °C to 37 °C in the presence or absence of the reducing agent TCEP. (C) Differential scanning calorimetry unfolding traces (*black*) of WT, W42Q, and W42R constructs with empirical three-state fittings (*red*) reveal the mutated N-terminal domains have reduced thermostability, but still melt above physiological temperatures. (D) Redox potential of the solution was set using the glutathione couple such that [GSH] + 2[GSSG] = 2 mM. Proteins were at 40 μM and 37 °C. Legend on right shows approximate OxD ("oxidation degree") values based on glutathione ratios *in vivo*, calculated from the data in (41). OxD = 2[GSSG]/(2[GSSG]+[GSH]). For example, OxD = 0.18 corresponds to [GSH]:[GSSG] ratio of 9:1.

**Figure 2: Lack of seeding effect in W42R aggregation**  180 μM W42R sample in the middle of its linear range of turbidity development was rapidly mixed 1:10 with fresh 180 μM sample of W42R (the mixing is indicated by the red arrow). No significant change in lag time or aggregation rate was observed upon seeding.

**Figure 3: Oxidation causes conformational change in W42Q and W42R**  4 μM W42Q (A), W42R (B), and WT (C) proteins were incubated for 60 minutes at 37 or 42 °C and OxD = 0.14 or 0.39. The protein concentration is below the aggregation range (~10% that in Figure 1). Fluorescence traces of the mutant proteins, but not the WT, show a red shift, indicating conformational change, particularly at higher temperature and oxidation. Heating these samples at the same temperatures in the absence of oxidizing agent did not result in any peak shift.

**Figure 4: RMSD from native structure for N- and C-terminal domains.** RMSD was averaged over 50 simulations at step 60,000,000. Grey line shows the midpoint of the sigmoidal fit to the WT curve. (A) N-terminal domain. (B) C-terminal domain. (C) Representative W42R structures populated at T = 0.800 with N-terminal domains in blue and C-terminal in red. Cys32 and Cys41 are shown as space filling models.

**Figure 5: Examples of internally rearranged W42R structures.** N-terminal domain is *blue*; C-terminal domain is *red*. (A) Rearrangement to enable binding of strand 1 to strand 8 within the N-terminal domain. (B) Rearrangement to enable binding of strand 1 to strand 14. Structures at simulated T = 0.8. Cys32 and Cys41 are shown as space-filling models.

**Figure 6: Protein-protein interactions from 2-molecule simulations of γD crystallin and predicted structure of aggregates.** (A) Initial simulation structure and final frame from a simulation trajectory, showing the N-terminal hairpin of one molecule binding to the folded C-terminal domain of the other molecule via interactions between strand 1 of the first molecule and the second strand of the C-terminal Greek key of the second molecule. (B) Percent of simulation frames showing protein-protein interactions, averaged over 300 trajectories. (C-G) Representative classes of interacting structures. (C) A native-like interaction, typical of the 1-4 disulfide. (D) A parallel strand 1 – strand 1 interaction, typical of the 1-2 disulfide. (E) An antiparallel strand 1 – strand 1 interaction, common for 1-2 and 2-3 disulfide. (F) An antiparallel strand 1 – strand 14 interaction, typical of the 2-3 disulfide. (G) A representative structure from the "Other" category. N-terminal domains are *blue*; C-terminal domains, *red*.

**Figure 7: Cys32-Cys41 is the predominant internal disulfide enabling W42Q aggregation.** (A) HPLC traces of aggregated (OxD = 0.2) and control (non-oxidized) W42Q samples following trypsin digestion under non-reducing conditions. Assignments for selected peaks, using electrospray mass spectrometry, are shown at right. (B, C, D) Oxidative aggregation of W42Q is highly temperature dependent in assays carried out as in **Figure 1**. (E) Cys mutations in the W42Q N-terminal domain alter its thermostability. Fitting for W42Q DSC unfolding shown in red (*dashed*) from Figure 1c is shown for comparison. Temperatures for the OxD aggregation assays on the C18A/W42Q (F) and C32V/W42Q (G) constructs were chosen to control for the differential thermostability. The former mutant aggregates already at 32 °C, while the latter does not do so even at 42 °C.

**Figure 8: Strand-strand contact maps from tethered two-chain W42R simulations involving the 2-3 disulfide.** (A) Most commonly interacting β-strands are strand 1 and the two strands most involved in the domain interface. (B) Combined contact maps for all simulations showing frequency of pairwise non-native strand-strand interactions.

**Figure 9**: A graphical model of aggregation by β-sheet completion between strand 1, in the N-terminal β-hairpin extruded due to formation of the Cys32-Cys41 internal disulfide bridge, and strand 14 in the C-terminal domain.

**TABLES**

| Sample | Tm(N), °C | Tm(C), °C | n |
|---|---|---|---|
| WT | 82.0 ± 0.4 | 84.9 ± 0.3 | 3 |
| W42Q | 54.6 ± 0.3 | 71.1 ± 1.7 | 2 |
| W42R | 56.1 ± 0.7 | 70.4 ± 0.5 | 5 |
| C18A/W42Q | 51.7 ± 0.2 | 74.1 ± 0.2 | 3 |
| C32V/W42Q | 60.2 ± 0.2 | 73.0 ± 0.1 | 3 |

**Table 1:** Melting temperatures of the N-terminal and C-terminal domains of the WT and several mutants of HγD. Data are reported as mean ± SD; *n* is the number of independent replicates.

**FIGURES**

**Figure 1**

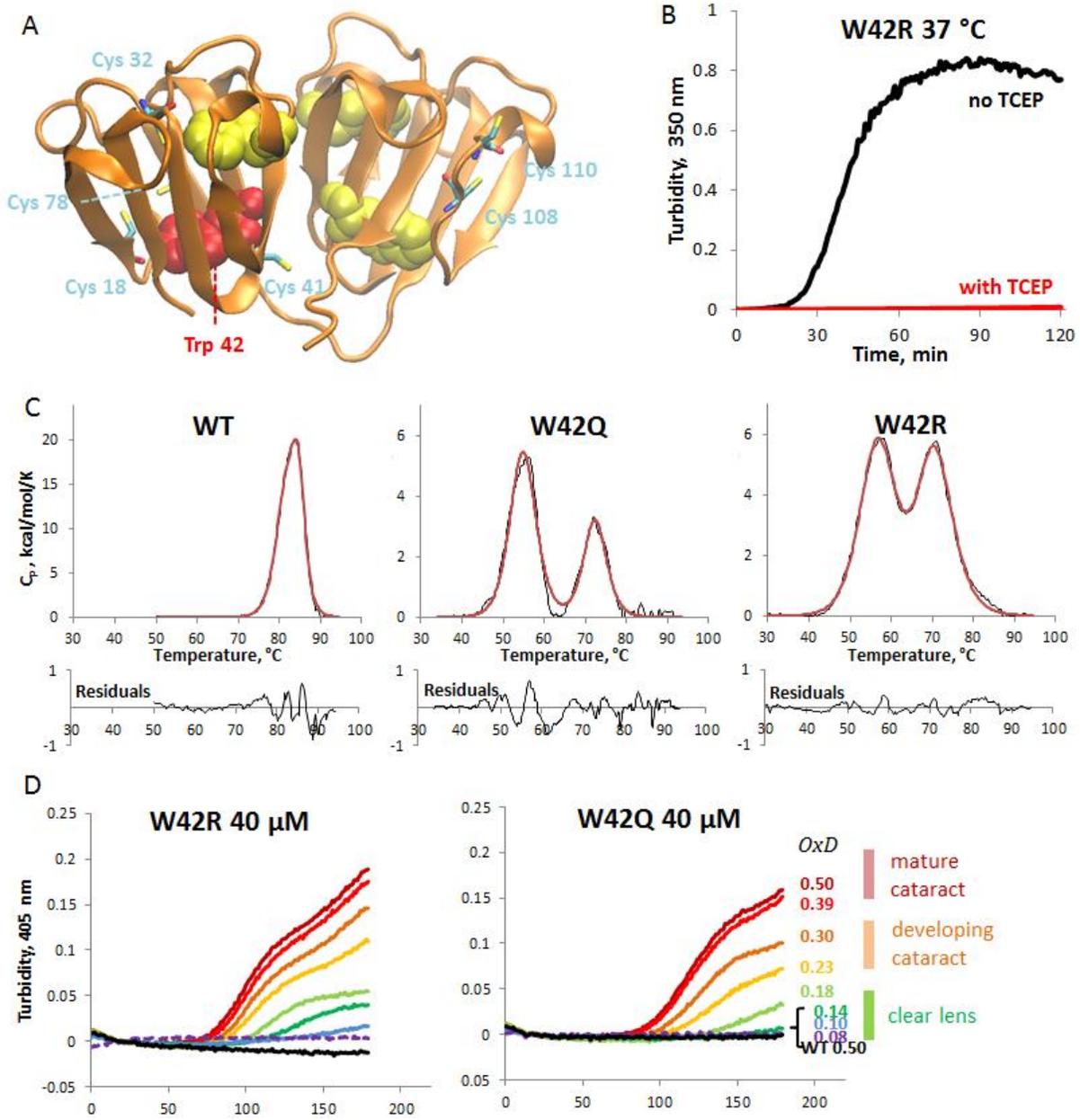

**Figure 2**

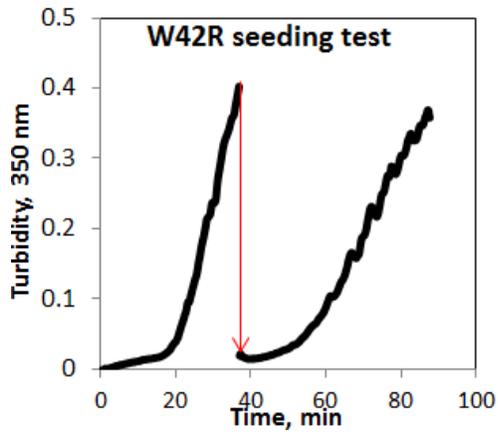

**Figure 3**

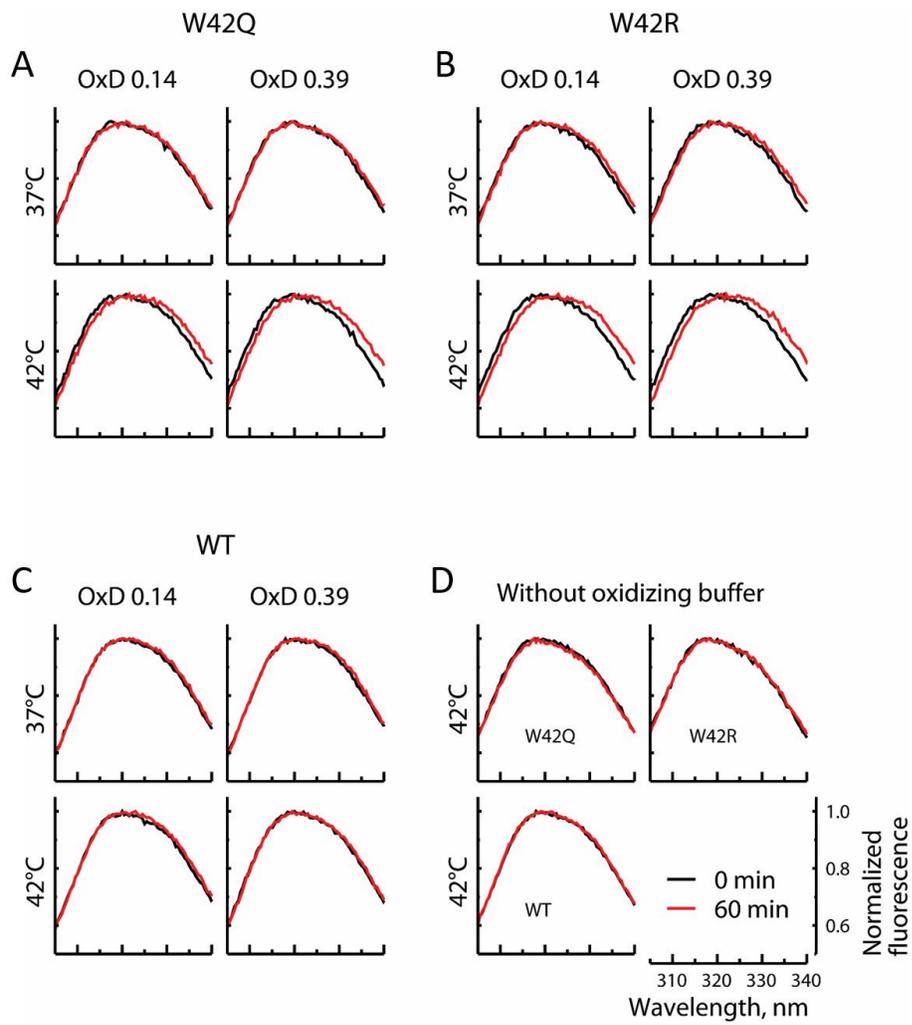

**Figure 4**

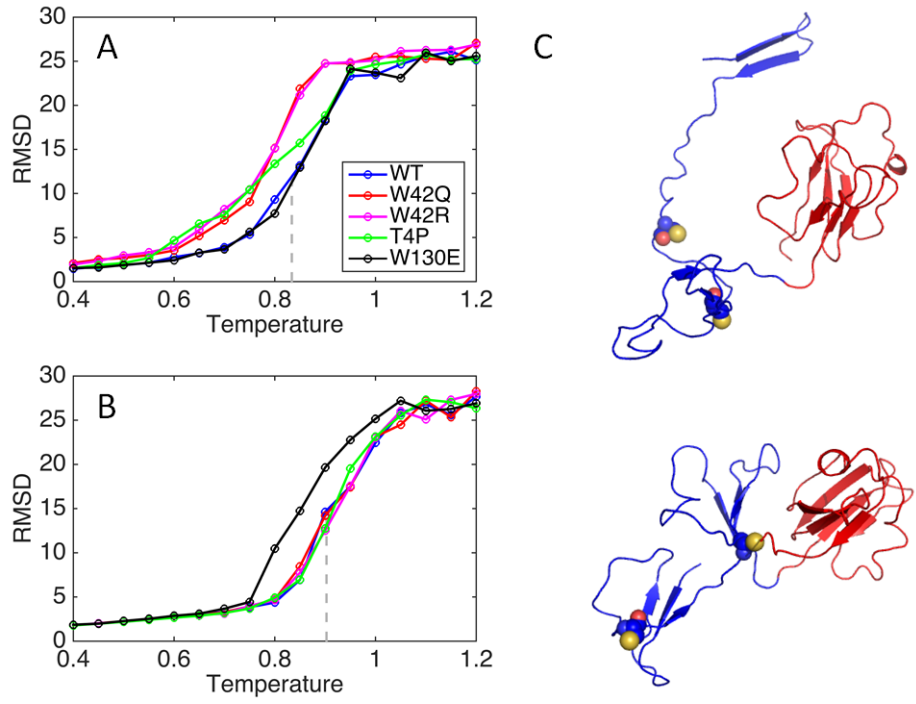

**Figure 5**

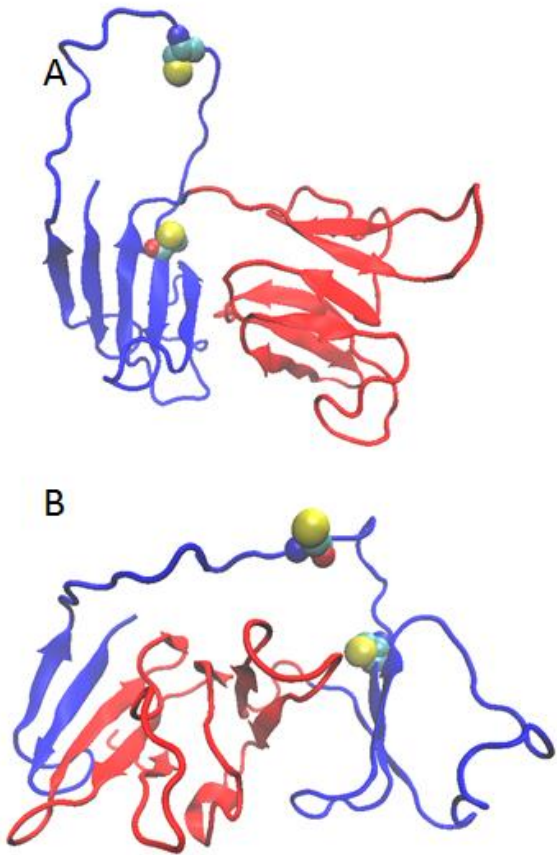

**Figure 6**

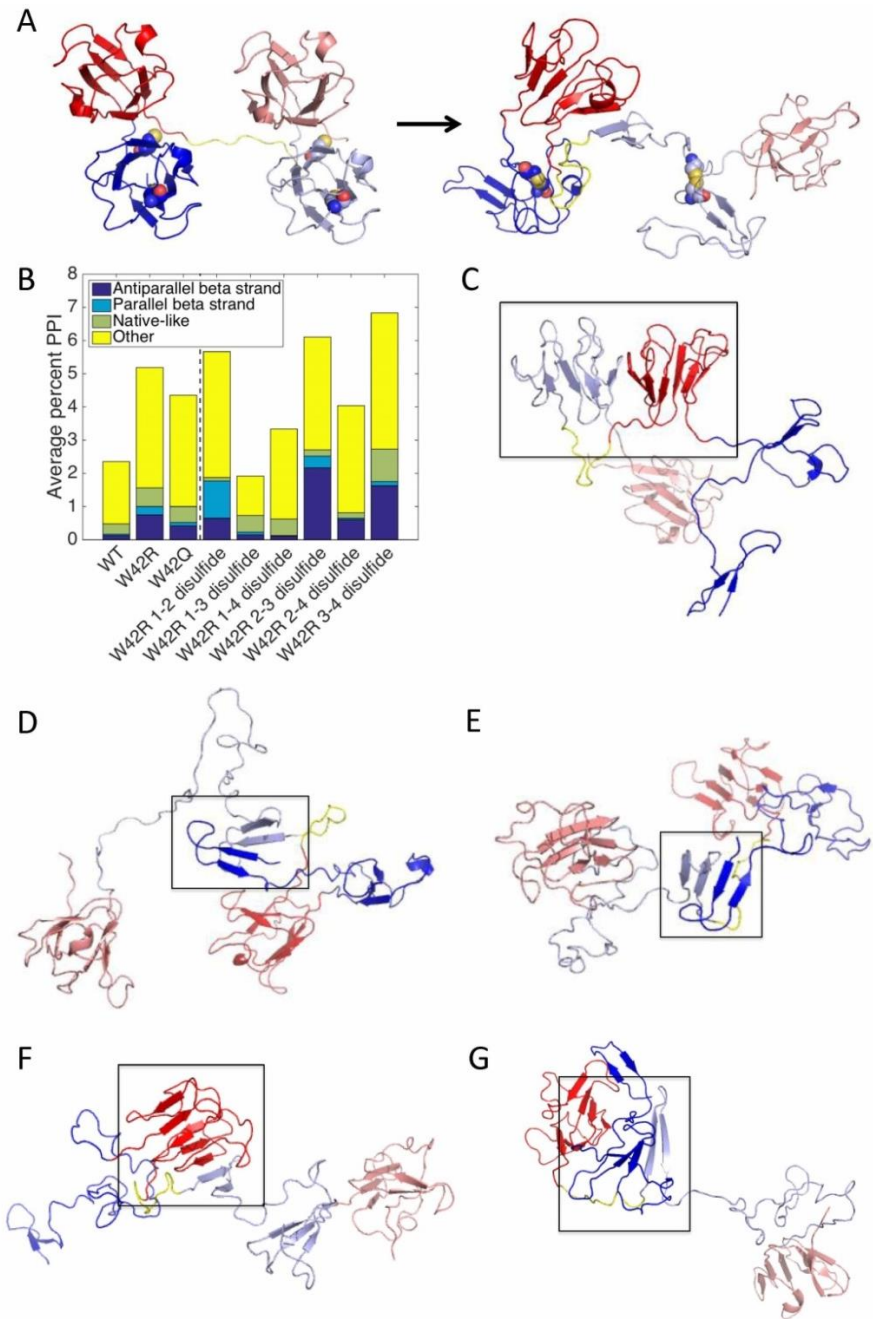

Figure 7

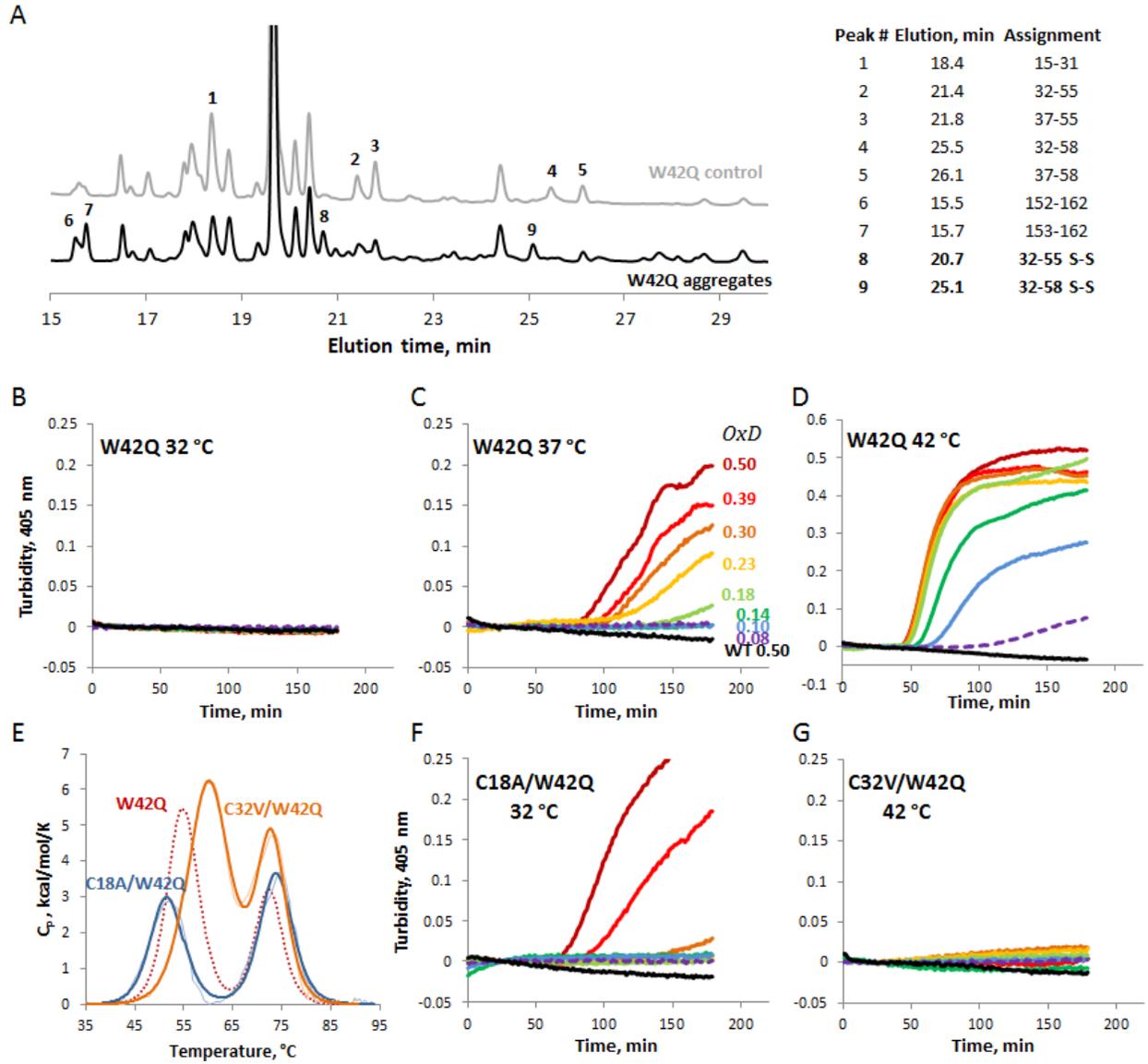

**Figure 8**

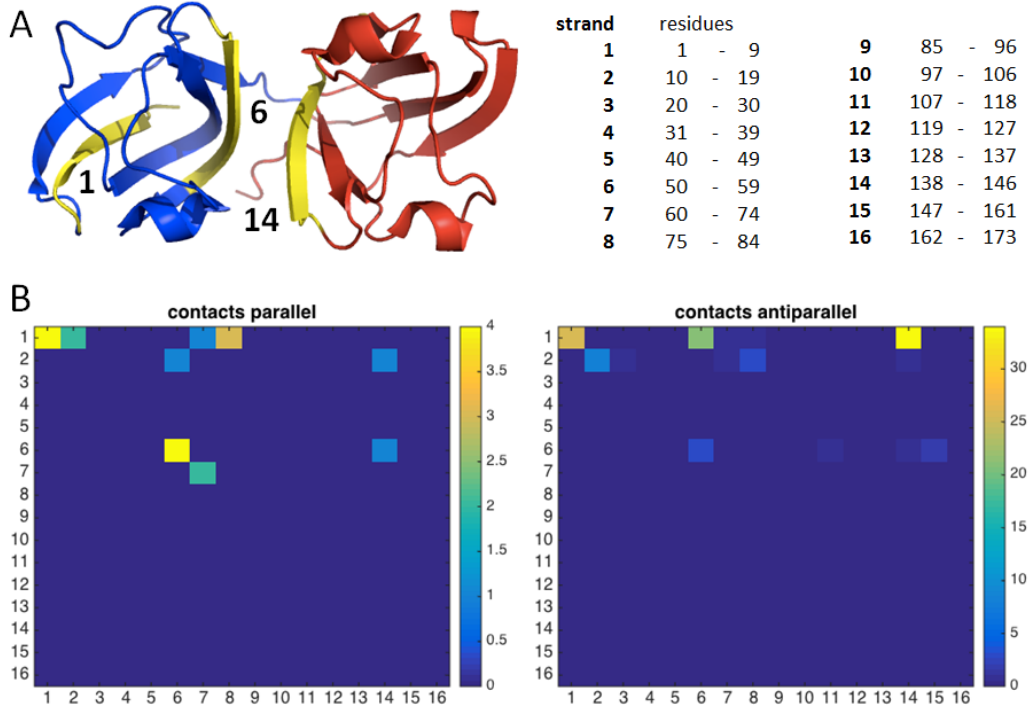

**Figure 9**

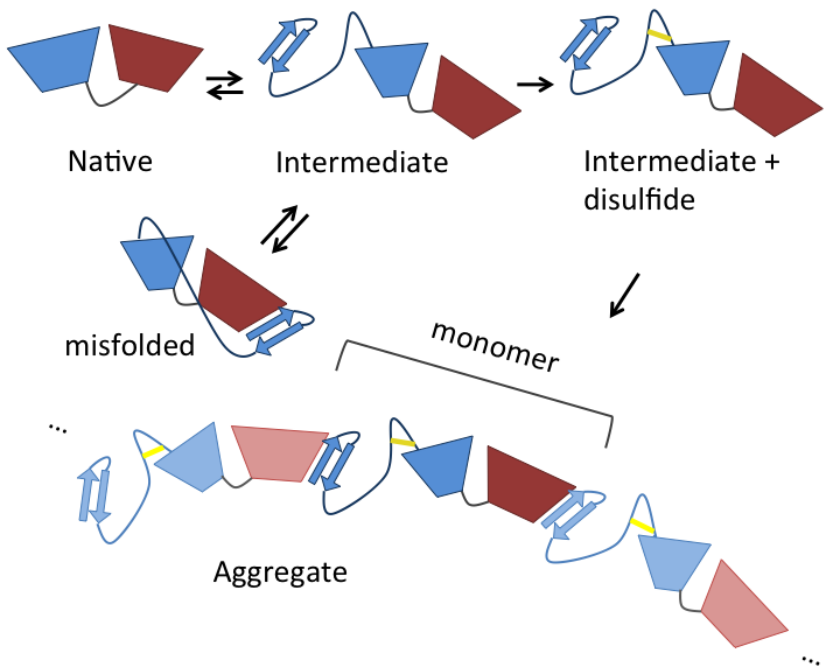